\documentclass[reprint,amsmath,amssymb,aps,prb,groupedaddress,nofootinbib,nobibnotes, floatfix, superscriptaddress,longbibliography]{revtex4-1}

\usepackage{optidef}
\usepackage{amsmath, amssymb, amsthm}
\usepackage{blkarray}
\usepackage{graphicx}
\usepackage{bm,bbm}
\usepackage{hyperref}
\usepackage[margin=2cm]{geometry}
\usepackage{xcolor}
\usepackage{algorithm}
\usepackage[noend]{algpseudocode}
\usepackage{multirow}
\theoremstyle{plain}
\usepackage{url}
\usepackage{mathtools}
\usepackage{float}
\usepackage{dcolumn}

\DeclareMathOperator*{\argmin}{arg\,min}

\DeclareMathOperator*{\Sp}{span}
\DeclareMathOperator*{\sgn}{sgn}
\DeclareMathOperator*{\Unif}{Unif}
\DeclareMathOperator*{\Adj}{Adj}

\algnewcommand\algorithmicforeach{\textbf{for each}}
\algdef{S}[FOR]{ForEach}[1]{\algorithmicforeach\ #1\ \algorithmicdo}

\begin{document}

\title{Phases of two-dimensional spinless lattice fermions with first-quantized deep neural-network quantum states}

\author{James Stokes$^{\star}$}
\email{jstokes@flatironinstitute.org}
\affiliation{
Center for Computational Quantum Physics, Flatiron Institute, New York, NY 10010 USA
}
\affiliation{
Center for Computational Mathematics, Flatiron Institute, New York, NY 10010 USA}

\author{Javier Robledo Moreno$^{\star}$}
\email{jrm874@nyu.edu}
\affiliation{
Center for Computational Quantum Physics, Flatiron Institute, New York, NY 10010 USA
}
\affiliation{Center for Quantum Phenomena, Department of Physics, New York University, 726 Broadway, New York, New York 10003, USA \\
$^{\star}$These authors contributed equally
}

\author{Eftychios A. Pnevmatikakis}
\email{epnevmatikakis@flatironinstitute.org}
\affiliation{
Center for Computational Mathematics, Flatiron Institute, New York, NY 10010 USA}

\author{Giuseppe Carleo}
\email{gcarleo@flatironinstitute.org}
\affiliation{
Center for Computational Quantum Physics, Flatiron Institute, New York, NY 10010 USA
}

\date{\today}

\begin{abstract}
First-quantized deep neural network techniques are developed for analyzing strongly coupled fermionic systems on the lattice. Using a Slater-Jastrow inspired ansatz which exploits deep residual networks with convolutional residual blocks, we approximately determine the ground state of spinless fermions on a square lattice with nearest-neighbor interactions.
The flexibility of the neural-network ansatz results in a high level of accuracy when compared to exact diagonalization results on small systems, both for energy and correlation functions. On large systems, we obtain accurate estimates of the boundaries between metallic and charge ordered phases as a function of the interaction strength and the particle density.
\end{abstract}

\maketitle

\section{Introduction}

The difficulty in treating interacting quantum systems stems directly from the fact that the state space of a many-body quantum system grows exponentially with the number of its constituents. The quantum many-body problem is a severe bottleneck in the understanding of complex quantum phenomena in many domains of science where quantum effects are relevant. When many-body effects are dominant, variational methods have proven a successful strategy to approximately represent, in a compact and computationally manageable form, many-body quantum states.

The prevailing paradigm for simulating lattice quantum systems in one spatial dimension is the density matrix renormalization group (DMRG) \cite{white_density_1992,white1993density}, which involves an iterative procedure to approximate low-entanglement quantum states using representations known as matrix product states\cite{fannes_finitely_1992,Perez-GarciaVWC07}. The success of DMRG to produce high overlap with the ground space stems from the ability of matrix product states to approximate gapped one-dimensional quantum systems\cite{hastings_area_2007} and the existence of a very efficient numerical scheme for their variational optimization. In two or more spatial dimensions, however, the situation is qualitatively very different and research into both computationally efficient and compact variational representations of quantum ground-states is very active.

A different context which directly confronts the curse-of-dimensionality is Machine Learning, that has recently found several applications in physics problems where high-dimensional functions are to be approximated\cite{carleo_machine_2019}.
The ability of generative models based on neural networks to overcome the curse-of-dimensionality in a variety of learning problems has motivated the development of neural network quantum states (NQS) and an associated real/imaginary-time evolution algorithm \cite{carleo2017solving} that extends the scope of the variational Monte Carlo \cite{mcmillan_ground_1965} to a number of challenging two-dimensional lattice systems.

Neural-network-based variational simulation has predominantly focused on systems corresponding to strongly localized electrons in which all spatial degrees of freedom have been frozen out, leaving an effective lattice Hamiltonian governing the spin degrees of freedom~\cite{carleo2017solving, Choo2019_J1-J2,ferrari_neural_2019,nomura_dirac-type_2020,vieijra_restricted_2020}.

Neural networks have also recently been proposed for simulating fermionic systems in the second quantized formalism \cite{choo2020fermionic}. This approach involves mapping the fermionic modes to an interacting quantum spin model, for example through a Jordan-Wigner transformation. The reduction of the fermionic Hamiltonian to a spin model makes it possible to capitalize on the successes of NQS for spin systems, but suffers from the disadvantage that the resulting spin Hamiltonian is nonlocally interacting. First quantization is an attractive alternative formalism, which preserves the locality of the physical interactions. In first quantization the solution of the quantum fermionic many-body problem can be posed as a function approximation problem, in which the target function to be approximated is a totally anti-symmetric solution of the time-independent Schrodinger equation.
First quantization has been explored for spinful Hubbard Hamiltonians, predominantly focusing on restricted Boltzmann machines \cite{nomura2017restricted}, Pfaffian states\cite{misawa2019mvmc} and backflow transformations\cite{luo2019backflow}. This approach has also been applied to ab-initio calculations of interacting electrons in the continuum~\cite{pfau2019DeepMind,hermann_deep_2019}. In this paper, we focus on the problem of approximating the ground-state for a model of two-dimensional spinless fermions with nearest-neighbor interactions, modeling the wavefunction using a Slater-Jastrow inspired factorization, with an additional neural network trained to capture sign deviations \cite{westerhout2020generalization} compared to the Slater determinant.

The paper is organized as follows: we begin by introducing the Hamiltonian, the qualitative features of the phase structure, and the observables that have been considered for identifying phase boundaries, as well as our proposed order parameter. We then discuss the detailed optimization problem and the relationship between variational Monte Carlo and other variational methods such as Hartree-Fock. Finally we present results for the phase structure and ground-state correlation functions.

\section{Theory}
\subsection{States and Hamiltonian}
Consider a system of spinless fermions hopping on the edges of a simple undirected graph $G=(\mathcal{V},\mathcal{E})$ with vertices $\mathcal{V}$ and edges $\mathcal{E}$. By Fermi statistics the number of fermions is constrained by $ N \in \{0,1 \ldots,|\mathcal{V}|\}$. For a $d$-dimensional hypercubic lattice with periodically identified boundaries, we have  $|\mathcal{V}|=L^d$ where $L$ is the side length of the lattice. The particle density is denoted by $\bar{n} = N/|\mathcal{V}|$. Let $\mathcal{V}^N$ be the set of $N$-tuples $x = (i_1,\ldots,i_N) \in \mathcal{V}^N$ and let $\mathcal{A}_N$ denote the complex vector space of totally antisymmetric functions mapping $\mathcal{V}^N \to \mathbb{C}$, which is of dimension $\dim_\mathbb{C} \mathcal{A}_N = C^{|\mathcal{V}|}_N$.

In order to describe the action of the Hamiltonian on the Hilbert space of states, we employ the following Fock space construction reviewed in the appendices. Fix an ordering $\leqslant$ on the vertices $i \in \mathcal{V}$. Then an orthonormal basis of $N$-particle states is given by $|x\rangle := \hat{c}^\dag_{i_1}\cdots \hat{c}^\dag_{i_N}|0\rangle$ with the $N$-tuples $x=(i_1,\ldots,i_N)$ restricted by the condition $i_1 < \cdots < i_N$ and where $|0\rangle$ denotes the Fock vacuum, which is annihilated by all $\hat{c}_i$.
The associated Hilbert space 
$\mathcal{H}_N=\Sp_\mathbb{C}\{|x\rangle {\; : \;} i_1 < \cdots < i_N \} $ is isomorphic to $\mathcal{A}_N$. A general state vector $|\Psi\rangle \in \mathcal{H}_N$ can be expanded over the basis as
\begin{align}
	|\Psi\rangle
		& = \sum_{i_1 < \cdots < i_N} \langle x | \Psi \rangle |x\rangle
		=: \sum_{x \in \mathcal{V}^N}f(x) |x\rangle \enspace ,
\end{align}
where in the second equality we have used the Fermi algebra to expand $|\Psi\rangle$ over a spanning set, with coefficients given by the output of the antisymmetric function $f \in \mathcal{A}_N$ defined by $f(x):=(1/N!) \langle  x | \Psi \rangle$. Conversely, any anti-symmetric function $f \in \mathcal{A}_N$ gives rise to a valid state vector. This paper focuses on variational families of states in which
the antisymmetric function $f \in \mathcal{A}_N$ is modeled using a  parametrized neural network.

The hopping dynamics on the graph $G=(\mathcal{V},\mathcal{E})$ is parametrized in terms of a pair of coupling parameters $t,V \geq 0$, which represent the kinetic energy and repulsive interaction strength, respectively,
\begin{equation}\label{Eq: Hamiltonian}
	\hat{H} = \sum_{\{i , j\} \in \mathcal{E}} -t(\hat{c}_i^\dag \hat{c}_j + \hat{c}_j^\dag \hat{c}_i) + V\hat{n}_i \hat{n}_j,
\end{equation}
where $\hat{n}_i = \hat{c}_i^\dag\hat{c}_i$. The above Hamiltonian commutes with the fermion number operator $\hat{N}=\sum_{i\in\mathcal{V}} \hat{n}_i$, making it possible to restrict to the fixed particle number subspace $\mathcal{H}_N$. 

This model Hamiltonian of spinless fermions in two dimensions is approximately realized in several systems of physical interest, including adsorbed submonolayers of spin-polarized $^3\textrm{He}_{\downarrow}$ and $D_{\downarrow}$~\cite{gubernatis1985lattice_gas, scalpino1948lattice_gas}, several organic materials (at one-quarter filling)~\cite{McKenzie2001RealMat} or ultracold atomic gases of spin-polarized $^6\textrm{Li}$~\cite{Moreo2007RealMat, Partridge2006RealMat}. The model is believed to exhibit a non-trivial phase transition between metallic and charge ordered phases. Despite its apparent simplicity, however, the phase diagram cannot be determined with high precision using existing numerical approaches and several questions remain open. These
include the nature of the charge ordered phase~\cite{Czart2008HF_phase_diag, Woul2010_pahse_diag} and the precise position of the phase boundary~\cite{gubernatis1985lattice_gas, scalpino1948lattice_gas, Czart2008HF_phase_diag, Woul2010_pahse_diag, song2014VMCspinless, sikora2015VMCspinless2}. Several state of the art variational wavefunctions have been applied to solve this model including the so called string-bond states~\cite{song2014VMCspinless} and tensor-product projected states~\cite{sikora2015VMCspinless2} (at half filling), as well as fermionic projected entangled-pair states (IPEPS)~\cite{corboz2010IPEPS} at arbitrary filling in the grand canonical ensemble. In this work we use neural network states to systematically investigate the phase diagram in the fixed particle subspace by considering a large collection of model parameters.

\begin{figure*}
        \includegraphics[width=.9\linewidth]{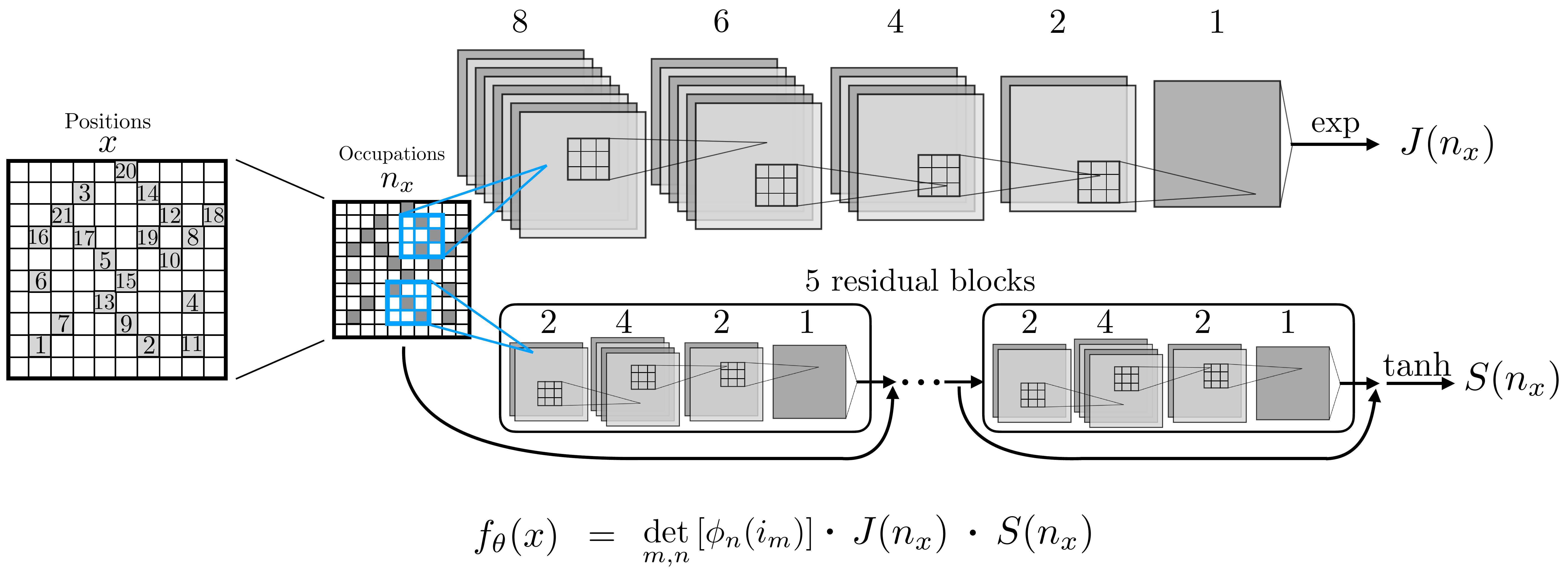}
        \caption{\label{FIG_00: network diagram}Diagram of the wavefunction ansatz as defined in Eq.~\ref{Eq: wavefunction ansatz}. From left to right: square lattice of side length $L = 10$ with $20$ particles, occupation map in the lattice that the sign and amplitude networks take as an input. Top and bottom convolutional networks are amplitude and sign factors respectively. The number of convolutional filters is indicated above each layer.}
\end{figure*} 

\subsection{Phase structure} 
Throughout this section, we focus exclusively on the $L \times L$ square with periodic boundary conditions. Moreover, we only consider even values of the side length $L$ such that $L^2$ is even so that the lattice supports half occupation. A rough picture of the $\bar{n}$-$V/t$ phase diagram can be determined by considering the limits of strong coupling ($V/t \to \infty$) and weak coupling ($V/t \to 0$), in which the model becomes exactly soluble for any value of $\bar{n}$. In the strong coupling limit $V/t\rightarrow \infty$, the system behaves like a hard-core classical lattice gas, in which translation invariance is broken as a result of charge-ordering. In the particular case of half occupation ($\bar{n}=0.5$) the charge-ordering is a staggered, checkerboard pattern and the corresponding phase is insulating. The non-interacting limit $V/t \to 0$ exhibits uniform density distribution and power-law density-density correlation functions, characteristic of a metallic phase. The above phase structure is qualitatively similar in one spatial dimensions, where the phase diagram can be exactly computed using the Bethe ansatz and bosonization\cite{Giamarchi2003}.

Although the phase diagram has been investigated using a variety of techniques~\cite{Wang2014HoneycombLattice, gubernatis1985lattice_gas, sikora2015VMCspinless2, corboz2010IPEPS,scalpino1948lattice_gas,song2014VMCspinless, Woul2010_pahse_diag,Czart2008HF_phase_diag}, there still exist open questions concerning the precise location of the phase boundary, (particularly at half occupation~\cite{sikora2015VMCspinless2, song2014VMCspinless, gubernatis1985lattice_gas, scalpino1948lattice_gas}) as well as the nature of the charge-ordered phase in the vicinity of the critical point~\cite{Woul2010_pahse_diag}.
 
An unrestricted Hartree-Fock (HF) analysis\cite{Czart2008HF_phase_diag} found that for sufficiently small values of the interaction, the system is in a gapless metallic state with uniform charge distribution. A critical value of $V/t$ was found, above which a first order phase transition leads to a thermodynamically unstable phase separation state where the system is comprised of both metallic and staggered charge-density wave components. More recently~\cite{Woul2010_pahse_diag}, an analysis using mean-field antinodal fermions refuted the hypothesis of a phase separation state, which was claimed to be an artifact of lack of accuracy of the unrestricted Hartee-Fock approximation, in a region of the phase diagram where the system is highly degenerate. They argue that the true nature of this state is a gapped, symmetry broken charge density wave (CDW) state with commensurate (staggered) charge order at half occupation and incommensurate charge order otherwise. Both of these analyses are formulated in the grand canonical ensemble, where the Hilbert space is chosen be the unrestricted Fock space, and the Hamiltonian is parametrized by a chemical potential $\mu \in \mathbb{R}$ as $\hat{H}(\mu) = \hat{H} -\mu \hat{N}$. The phase diagram was constructed by identifying cusps in the ground state energy of $\hat{H}(\mu)$ as a function of $\mu$. In contrast, since we restrict to a subspace $\mathcal{H}_N$ defined by fixed particle number, the chemical potential term only contributes an irrelevant constant to the energy, so we require a different strategy to find the transition points. 

 Different observables have been considered for detecting the phase transition at fixed particle number. In particular, at half-occupation ($\bar{n} = 0.5$), the phase transition from a metallic phase to a checkerboard charge-ordered insulating phase is detected by the so-called charge structure factor $S(\pi,\pi)$, defined as the $\bm{k}=(\pi,\pi)$ component of the Fourier transform of the two-point correlation function averaged over lattice locations~\cite{gubernatis1985lattice_gas, sikora2015VMCspinless2, scalpino1948lattice_gas, song2014VMCspinless}.

This observable abruptly increases upon crossing to the charge ordering phase due to the staggered charge ordering~\cite{gubernatis1985lattice_gas, scalpino1948lattice_gas}. The charge structure factor $S(\pi,\pi)$ is not suitable for identifying the transition away from half-filling because it assumes that the charge-ordering is commensurate with the underlying lattice.

In order to define an order parameter suitable for general filling fraction, we first identify the vertex set with the two-dimensional periodic torus $\mathcal{V} \cong \mathbb{Z}_L^2$ (where $\mathbb{Z}_L = \{0,\ldots,L-1\}$) and
 define a density signal $\rho : \mathbb{Z}_L^2 \to [0,1]$ on the torus by $\rho(\bm{r}) = \langle \hat{n}_{\bm{r}} \rangle$ where $\bm{r} \in \mathbb{Z}_L^2$. We seek an order parameter $o$ that measures the departure of this density signal from homogeneity so we define
\begin{equation}\label{Eq: order parameter}
    o[\rho] := \frac{L^2}{N(L^2-N)}\left[\Vert \rho \Vert_2^2 - \frac{1}{L^2} \Vert \rho \Vert_1^2\right] \geq 0 \enspace .
\end{equation}
Positivity of the order parameter follows directly from the relation between the $l_1$ and $l_2$ norms. The bound is saturated when $\rho$ is a constant function (uniform density) and the overall normalization is chosen such that the order parameter is unity for a classical $N$-particle state. Moreover, by Parseval's identity, this order parameter is related to the energy in the nonzero Fourier modes
\begin{equation}
    o[\rho] = \frac{1}{N(L^2-N)}\sum _{\bm{k}\neq \bm{0}} | \widetilde{\rho}(\bm{k})|^2 \enspace ,
\end{equation}
where the $\bm{k}$ sum is over all nonzero modes in the discrete Brillouin torus $\frac{2\pi}{L}\mathbb{Z}_L^2$. This quantity is evidently dependent on multiple Fourier modes, as required to capture incommensurate order. Therefore, in the thermodynamics limit, this observable vanishes in the metallic phase and becomes nonzero and finite upon the formation of a charge ordered state. 

Another observable which has received attention in this context is the so-called density-density correlation function, which is defined for\footnote{The domain of defintion of $C$ follows from the fact that $L$ is assumed to be even.} $r \in \mathbb{Z}_{L+1} = \{0, \ldots, L \}$ as follows,
\begin{equation}\label{Eq: Correlation definition}
C(r) = \frac{1}{L^2 N_r}\sum_{i \in \mathcal{V}} \sum_{j \in S_r(i)}\left\langle \left(\hat{n}_i-\bar{n}\right)\left( \hat{n}_{j} - \bar{n}\right) \right\rangle
\end{equation}
where $S_r(i) = \{j \in \mathcal{V} : d(i,j)=r\}$ is the set of vertices with graph distance $r$ from $i \in \mathcal{V}$ and $N_r = |S_r(i)|$ is the number of such vertices, which is constant for the square lattice under consideration.

Due to the expected short-distance divergences of the two-point correlation function in the continuum limit, we define the renormalized Fourier space correlator by subtracting the coincidence limit of the position space correlator
\begin{equation}\label{e:Cren}
    \widetilde{C}_{\rm ren}(k) 
    := \sum_{r=0}^L e^{- i k r} C_{\rm ren}(r)
    = \sum_{r=1}^L e^{- i k r} C(r)
\end{equation}
where $k \in \frac{2\pi}{L+1}\{0,\ldots, L\}$.

\section{Methods}
In this section we describe the variational Monte Carlo and Hartree-Fock optimization problems and our proposed variational ansatz.

\subsection{Optimization problem}
For each integer particle number $N$ in the range  $1 \leq N \leq |\mathcal{V}|$, we consider the problem of finding a minimal energy  simultaneous eigenvector $|\Psi_0\rangle$ of both the Hamiltonian $\hat{H}$ and the number operator $\hat{N}$ such that $\hat{N} | \Psi_0\rangle = N |\Psi_0\rangle$ and $\hat{H} |\Psi_0\rangle = E_0 |\Psi_0\rangle$. Such an eigenvector admits a characterization in terms of the Rayleigh quotient as follows,
\begin{equation}
	|\Psi_0\rangle \in \argmin_{\Psi \in \mathcal{H}_N : \Psi \neq 0} \frac{\langle \Psi | \hat{H} | \Psi \rangle}{\langle \Psi | \Psi \rangle} \enspace .
\end{equation}
In the case of the interacting Hamiltonian \eqref{Eq: Hamiltonian}, optimization over all $N$-particle wavefunctions is intractable, so we focus on the simpler problem of selecting a wavefunction from a variational class of trial wavefunctions.  A subset of $\mathcal{H}_N$ is chosen by specifying a family $\mathcal{F} \subseteq \mathcal{A}_N$ of antisymmetric functions. For each $f \in \mathcal{F}$, the associated wavefunction $|\Psi_f\rangle \in \mathcal{H}_N$ gives an upper bound on the ground-state energy,
\begin{equation}
    E_0 \leq  \frac{\langle \Psi_f | \hat{H} | \Psi_f \rangle}{\langle \Psi_f | \Psi_f \rangle} \enspace ,
\end{equation}
and thus optimizing this bound over $\mathcal{F}$ yields an approximation of the ground-state eigenpair.

By standard arguments reviewed in the appendices, the above Rayleigh quotient can be expressed as a classical expectation value of a local energy functional
\begin{equation}\label{e:E_f}
	\frac{\langle \Psi_f | \hat{H} | \Psi_f \rangle}{\langle \Psi_f | \Psi_f \rangle} =  \underset{x \sim \pi_f}{\mathbb{E}}[E_f(x)] \enspace ,
\end{equation}
where $\pi_f$ is a permutation-invariant probability distribution over the classical state space $\mathcal{V}^N$, which assigns a probability to $x \in \mathcal{V}^N$ given by,
\begin{equation}
	\pi_f(x) = \frac{|f(x)|^2}{\sum_{x' \in \mathcal{V}^N}|f(x')|^2}
\end{equation}
and the local energy $E_f$ is a permutation-invariant function of $x \in \mathcal{V}^N$ defined by
\begin{equation}
    E_f(x) = -t \sum_{x' \in \Delta_x} \frac{f (x')}{f (x)} + V\sum_{\{i,j\} \in \mathcal{E}} n_i n_j \enspace ,
\end{equation}
where $\Delta_x \subseteq \mathcal{V}^N$ denotes the set of classical states obtained by applying the kinetic operator to $|x\rangle$, and $n_i \in \{0,1\}$ denotes the binary occupation number of vertex $i \in \mathcal{V}$.
In practice, the family of antisymmetric functions $\mathcal{F}$ is parametrized by unconstrained variational parameters $\theta \in \mathbb{R}^d$, and we locally optimize the following loss function
\begin{equation}
    L(\theta) := \underset{x \sim \pi_{f_\theta}}{\mathbb{E}}[E_{f_\theta}(x)] \enspace .
\end{equation}

Sampling from the probability density $\pi_f$ was performed using a Markov chain Monte Carlo strategy outlined in Algorithm~\ref{alg: mcmc}. The configuration space of the Markov chain is given by the classical state space $\mathcal{V}^N$. It is convenient to maintain a lookup table $\kappa : \mathcal{V} \to [N] \cup \{\mathsf{empty}\}$ which returns the location of a given vertex in the array $x=(i_1,\ldots,i_N)$, or $\mathsf{empty}$ if absent. 

\begin{algorithm}[H]
\caption{Markov Chain Metropolis \label{alg: mcmc}}
\begin{algorithmic}[1]
\State Initialize $x = (i_1,\ldots, i_N) \in \mathcal{V}^N$
\For{$t =1 $ to $T$}
\State Sample particle $n \sim \Unif(\{1, \ldots, N\})$
\State Sample vertex $i \sim \Unif(\Adj(x_{t-1}(n)))$
\If{$\kappa_{t-1}(i) \neq \mathsf{empty}$}
\State $x_t = x_{t-1}$
\Else
\State $x' = x_{t-1}$
\State $i_n' \gets i$
\State $q \sim \Unif([0, 1])$
\If{$q < |f(x')|^2 / |f(x_{t-1})|^2$}
\State $x_t = x'$
\Else
\State $x_t = x_{t-1}$
\EndIf
\EndIf
\EndFor
\end{algorithmic}
\end{algorithm}

 \begin{figure}
        \includegraphics[width=1\linewidth]{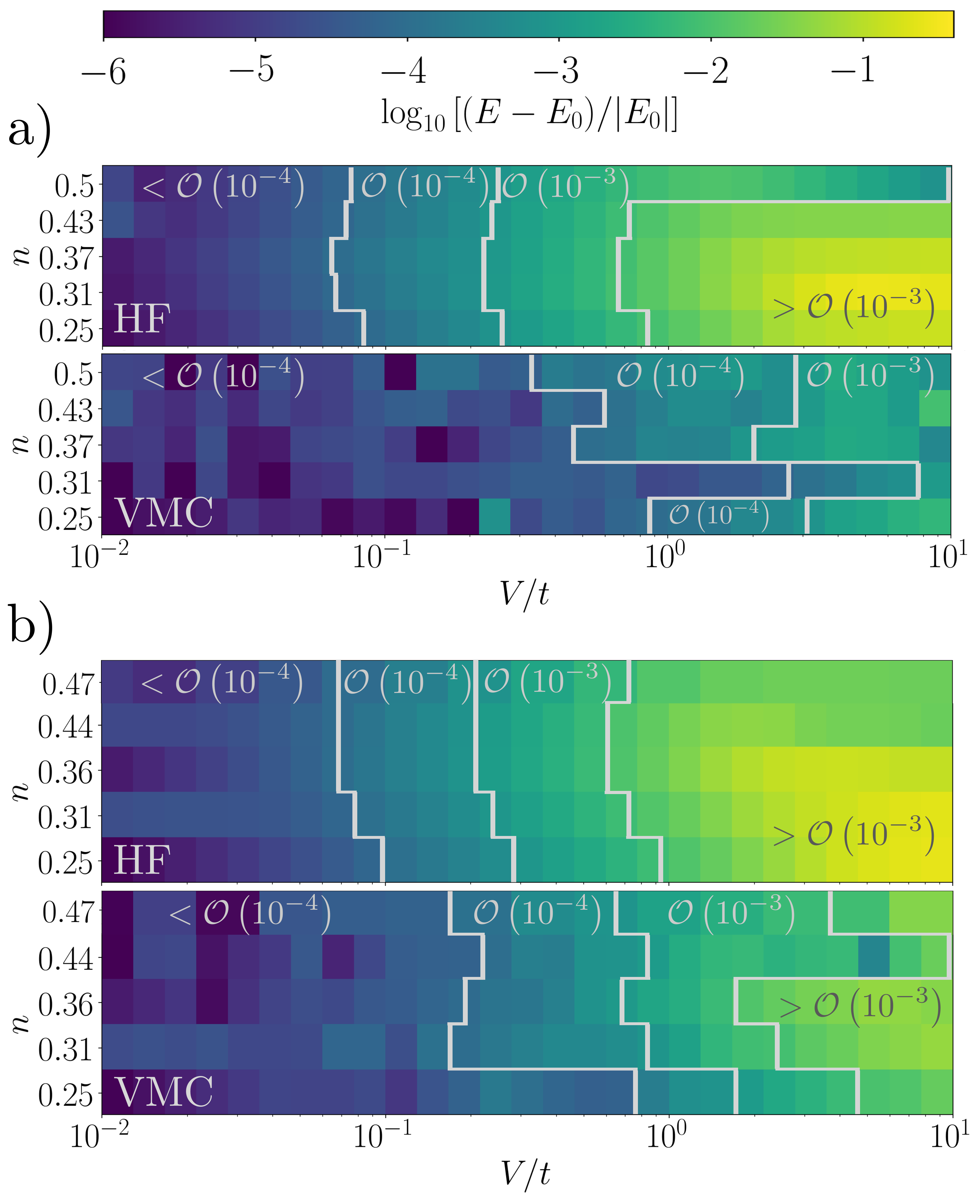}
        \caption{\label{FIG_01: ED Energy Benchmark} Benchmark of the proposed variational ansatz (VMC) and Hartree-Fock approximation (HF) using the exact ED states. Color-maps show the relative error of the energy in the square lattice of size $L = 4$  (panel \textbf{a)}) and in the square lattice of size $L = 6$ (panel \textbf{b)}). Relative error is shown at different fillings of the lattice and values of the coupling constant. Lines separate regions of the phase diagram with different orders of magnitude of the relative error as indicated.}
\end{figure} 
\begin{figure}
        \includegraphics[width=1\linewidth]{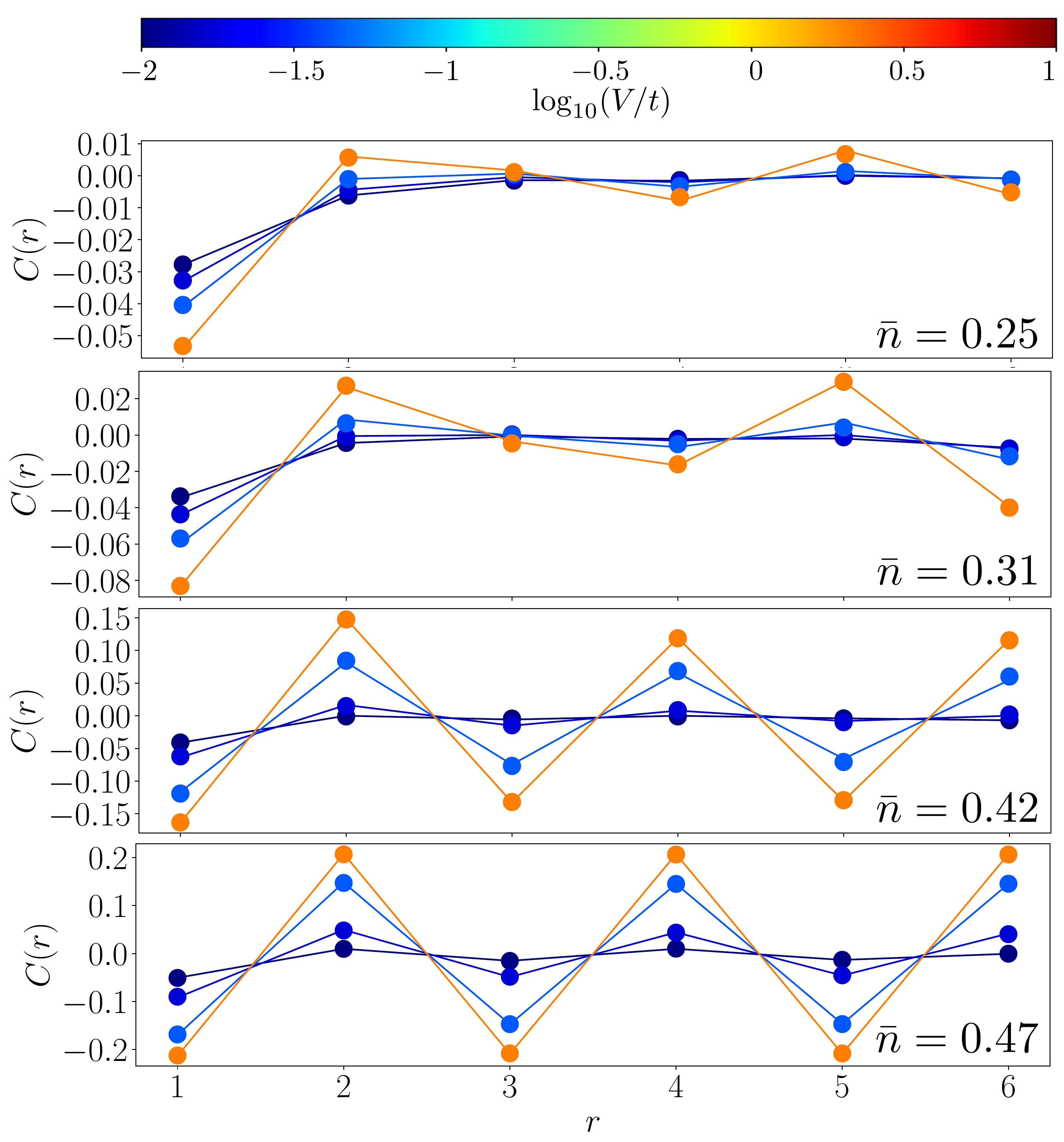}
        \caption{\label{FIG_02: ED corr Benchmark} Benchmark of the proposed variational ansatz (VMC) using ED eigenpairs in the square lattice of side length $L = 6$.  Two-point density correlation functions (Eq.~\eqref{Eq: Correlation definition}) are shown as a function of graph distance $r$  at different values of $V/t$: $V/t = 0.01$, $V/t = 0.599$, $V/t = 2.15$ and $V/t = 5.99$, as shown by the color scale in the top of the panel. Solid lines are the correlations from ED and the dots correspond to the correlations computed with our approach. Different panels correspond to different fillings as indicated in each panel.}
\end{figure}

\subsection{Hartree-Fock}
It is instructive to contrast the VMC optimization problem with Hartree-Fock, which provides one of our baselines.
Suppose that the matrix $\phi \in \mathbb{C}^{|\mathcal{V}| \times N}$ is an isometric matrix; that is, $\phi^\dag \phi =\mathbbm{1}_N$ and therefore $P = \phi \phi^\dag$ is a Hermitian projection onto the image of $\phi$. If we define a family of $N$ creation operators,
\begin{equation}
    \widetilde{c}_n^\dag = \sum_{i \in \mathcal{V}} \hat{c}_i^\dag \phi_n(i) \enspace ,
\end{equation}
and define the $N$-particle normalized Hartree-Fock state,
\begin{equation}
    |\Psi_{\rm HF}\rangle = \widetilde{c}_{1}^\dag \cdots \widetilde{c}_{N}^\dag |0\rangle,
\end{equation}
then it is easily shown that the entries of the projection matrix $P \in \mathbb{C}^{|\mathcal{V}| \times |\mathcal{V}|}$ are given by
\begin{equation}
    P_{ij} = \langle \Psi_{\rm HF} |\hat{c}_j^\dag \hat{c}_i| \Psi_{\rm HF} \rangle.
\end{equation}
Expressing the Hamiltonian in terms of the adjacency matrix $A$ for the graph,
\begin{equation}
	\hat{H} = \sum_{(i,j) \in \mathcal{V}^2} A_{ij} \left[ -t \hat{c}_i^\dag \hat{c}_j - \frac{1}{2}V \hat{c}_i^\dag \hat{c}_j^\dag \hat{c}_i \hat{c}_j \right],
\end{equation}
where we have used the fact that $A$ is zero on the diagonal.
The quantum expectation value of the Hamiltonian in the Hartree-Fock state is thus given by
\begin{align}
    \langle \Psi_{\rm HF} | \hat{H} | \Psi_{\rm HF} \rangle
        & = \sum_{(i,j) \in \mathcal{V}^2} A_{ij} \left[ -t P_{ji} -\frac{1}{2}V  (P_{ij}P_{ji}-  P_{ii}P_{jj}) \right]
\end{align}
The optimization problem is to minimize the energy $E_{\rm HF}(\phi) := \langle \Psi_{\rm HF} | \hat{H} | \Psi_{\rm HF} \rangle$ viewed as a function of the isometric matrix $\phi$,
\begin{mini}|l|
  {}{E_{\rm HF}(\phi)}{}{}
  \addConstraint{\phi^\dag \phi }{=\mathbbm{1}_N}{}
\end{mini}
In practice, we approximated the solution of the above constrained optimization problem by alternating between steps of Adam optimization followed by orthogonalization of the columns of $\phi$.

\subsection{Wavefunction ansatz}
Since the focus of this paper is ground-state optimization, we restrict to real-valued wavefunctions by exploiting time-reversal invariance of the Hamiltonian. In order to describe our choice of antisymmetric functions $\mathcal{F}$, it is useful to define an indicator vector $n_x \in \{\pm\}^{|\mathcal{V}|}$ for the configuration $x\in \mathcal{V}^N$, which is given in terms of the re-centered binary occupations
\begin{equation}
    n_x := (2n_i-1)_{i \in \mathcal{V}}
\end{equation}
where $n_i \in \{0,1\}$.
The family of antisymmetric functions $\mathcal{F}$ is chosen to consist of parametrized functions $f_\theta$ possessing a generalized Jastrow-Slater form, meaning that their dependence on the occupation numbers factorizes from a Slater determinant as follows
\begin{equation}\label{Eq: wavefunction ansatz}
    f_\theta(x) = \psi_0(x) \, J(n_x) \, S(n_x) \enspace .
\end{equation}
The domain and range of the constituent functions appearing in the above factorization is given as follows,
\begin{align}
    & \psi_0 : \mathcal{V}^N \to \mathbb{R} \\
    & J :\{\pm\}^{|\mathcal{V}|} \to [0,\infty) \\
    & S : \{\pm\}^{|\mathcal{V}|} \to [-1,1].
\end{align}
In particular, $\psi_0$ is a Slater determinant, while $J$ and $S$ are neural networks, chosen with the property that they are invariant under a subset of lattice symmetries, and furthermore that they approach constant functions for some setting of the parameters. The proposed ansatz can thus be viewed as a deformation of the Hartree-Fock wavefunction by a Jastrow factor and an additional factor that corrects the sign structure. The fact that $J$ and $S$ can exactly represent constant functions implies that the proposed ansatz is exact in the non-interacting ($V/t = 0$) and classical ($V/t \rightarrow \infty$) limits.

The variational parameters characterizing the Slater determinant consist of a $|\mathcal{V}| \times N$ matrix $\phi = \left[\phi_1,\ldots \phi_N\right] \in \mathbb{R}^{|\mathcal{V}| \times N}$. Denoting the $i$th entry of the $n$th column $\phi_n \in \mathbb{R}^{|\mathcal{V}|}$  by $\phi_n(i)$, it follows that the Slater determinant is
$\psi_0(x) = \det_{m,n} \left[ \phi_n(i_m)\right]$
where the orbital functions $\phi_n : \mathcal{V} \to \mathbb{R}$ are neither normalized, nor orthogonal.

The remaining variational parameters $\theta_J \in \mathbb{R}^{d_J}$ and $\theta_S \in \mathbb{R}^{d_S}$ characterize the weights and biases of the neural networks $J$ and $S$. In order to meet the desiderata of translational invariance and ability to represent the constant function, we choose $J$ to be a convolutional feed-forward network with output exponential nonlinearity and we chose $S$ to be a deep residual network \cite{he2015resnet} with convolutional residual blocks, followed by an averaging layer, a fixed affine transformation\footnote{In practice, we chose the affine layer to be pointwise application of $x\mapsto\frac{x}{2(2N-|\mathcal{V}|)} + \frac{1}{2}$.} and final output tanh nonlinearity. In both cases the convolutions employed periodic boundary conditions (see Fig.~\ref{FIG_00: network diagram}). 

\begin{figure*}
        \includegraphics[width=1\linewidth]{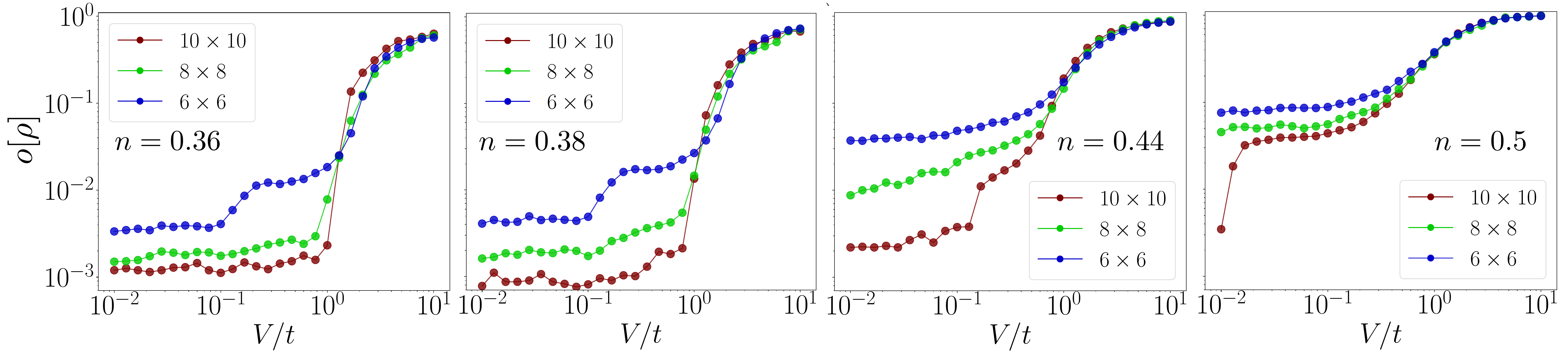}
        \caption{\label{FIG_05: density power spectrum} Order parameter as defined in Eq.~\eqref{Eq: order parameter} as a function of the coupling constant. Each panel shows the order parameter in system sizes $L = 6$, $L = 8$ and $L = 10$. as indicated in the legend with different colors. Lines connecting dots are for visual guidance. Different panels correspond to different fillings as indicated.} 
\end{figure*}

The variational parameters $\theta = (\phi, \theta_J,\theta_S) \in \mathbb{R}^d$ ($d = |\mathcal{V}| N + d_J + d_S$) were jointly optimized using the stochastic reconfiguration method~\cite{sorella2007StochasticReconfiguration}, which can be interpreted as imaginary time evolution or a particular case of the natural gradient optimization~\cite{amari1998natural, stokes2020quantum}. The variational parameters $(\theta_J,\theta_S)$ were intialized using standard random initalization strategies and the Slater determinant parameters $\phi$ were initialized using the solution of the Hartree-Fock optimization scheme described above for a faster convergence.
\begin{figure}[H]
        \includegraphics[width=.9\linewidth]{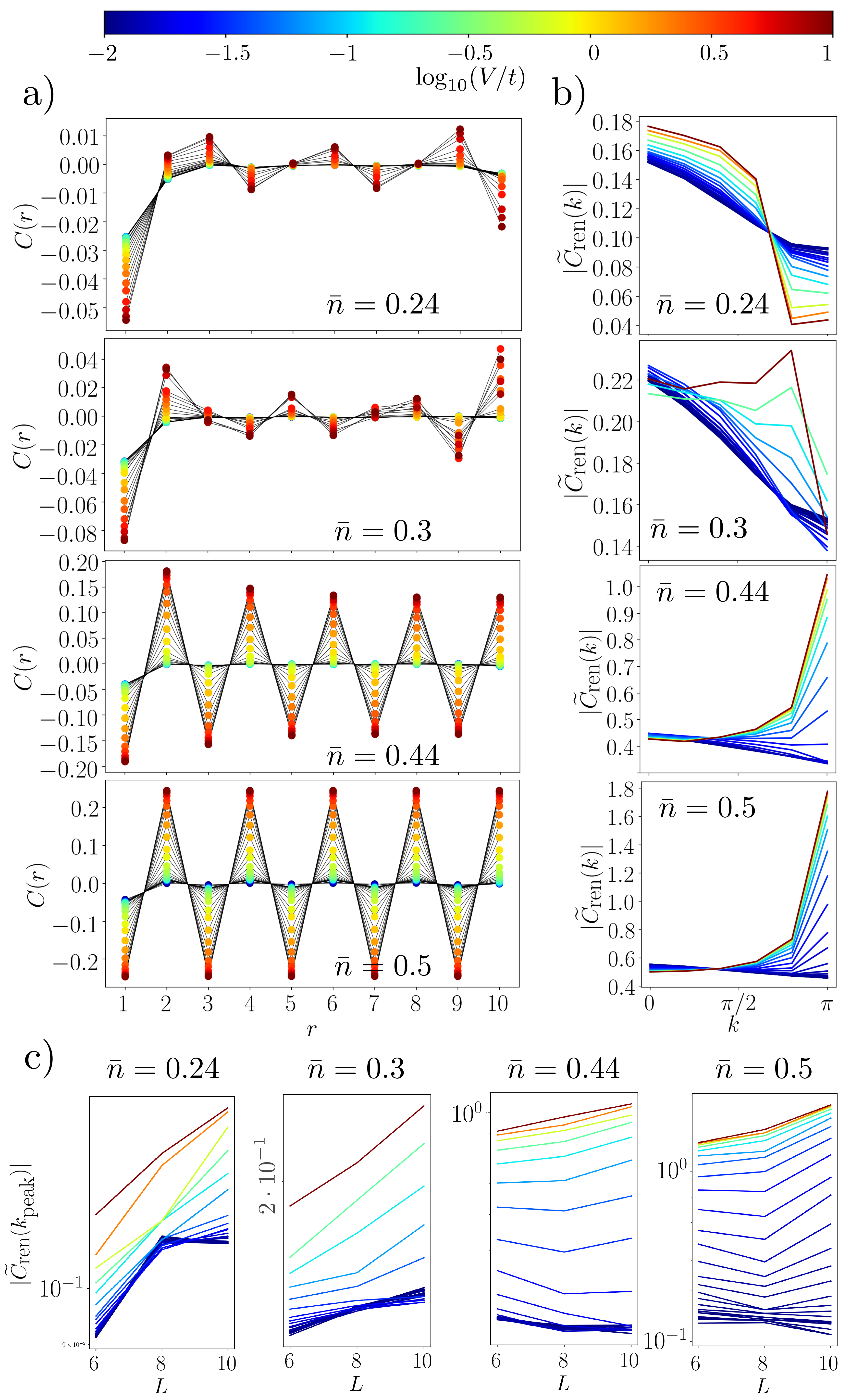}
        \caption{\label{FIG_03: radial correlations} Two-point density  correlation function defined in \eqref{Eq: Correlation definition} and the moduli of their renormalized Fourier modes defined in \eqref{e:Cren}. Different colors indicate different values of $V/t$ as indicated in the legend in the top of the plot. \textbf{a)} Two-point density correlation functions as a function of the graph distance $r$ . System size is $L = 10$. Dots represent the obtained values for the correlations and black solid lines are for visual guidance. Each panel corresponds to a different filling as indicated. \textbf{b)} Modulus of the Fourier transform of the correlation functions displayed in the corresponding a) panels, as a function of $k$. Note that $|\widetilde{C}_{\rm ren}(k)|$ is only shown for $k \in [0, \pi]$ as $|\widetilde{C}_{\rm ren}(k)|$ is symmetrical with respect to $k = \pi$. \textbf{c)} amplitude peak in $|\widetilde{C}_{\rm ren}(k)|$ shown in a), as a function of the lattice size. For different system sizes the peak is identified as the maximum of $|\widetilde{C}_{\rm ren}(k)|$ at the largest value of $V/t$ considered. Different panels correspond to different lattice fillings as indicated.}
\end{figure} 

\section{Results}
In this section we analyze the ground-state correlation functions and the phase structure by performing ground-state optimization of the neural network ansatz using variational Monte Carlo. In particular, two-point density correlation functions are analyzed to probe the nature of the charge-ordered phase. Then the phase diagram is constructed  by performing a finite-size scaling analysis of the density order parameter \eqref{Eq: order parameter}.

\subsection{Exact diagonalization benchmark}
In order to verify the implementation, the exact ground state eigenpair was determined using the Lanczos algorithm as implemented in QuSpin~\cite{Quspin2017, Quspin2019}, working in the zero momentum sector containing the ground state. Exact diagonalization is practical for system sizes up to $L=6$ with $N \leq 17$ particles.

The relative error in the
ground-state energy obtained using VMC optimization of the neural network is shown in Fig.~\ref{FIG_01: ED Energy Benchmark}, alongside the relative error obtained using unrestricted Hartree-Fock for comparison. Unsurprisingly, the neural network ansatz outperforms the unrestricted HF in terms of accuracy of ground-state energy, with relative errors smaller by up to two orders of magnitude for certain values of the coupling. For most values of $V/t$, the relative error does not exceed $\mathcal{O} (10^{-3})$. In the worst cases, which correspond to large values of the interaction (but still far from the classical limit), the relative error is no larger than $\mathcal{O} (10^{-2})$. 
The ground-state energies obtained using our method were found to be lower than those obtained using tensor-product projected states~\cite{sikora2015VMCspinless2} and string bond states~\cite{song2014VMCspinless}, which are available for system size $L=4$ and half occupation ($\bar{n}=0.5$).

In order to benchmark the ground-state wavefunction beyond the energy error, we computed density-density correlations as defined in Eq.~\eqref{Eq: Correlation definition} and compared against ED results, as shown in Fig.~\ref{FIG_02: ED corr Benchmark}. The neural network ansatz accurately reproduces the exact correlation functions for any graph distance $r$ at any $V/t$ and filling values. 

The benchmark with ED shows that the proposed neural network variational ansatz provides an accurate approximation to the ground-state, making it a suitable tool to study the phase diagram of the model.

\begin{figure*}
        \includegraphics[width=1\linewidth]{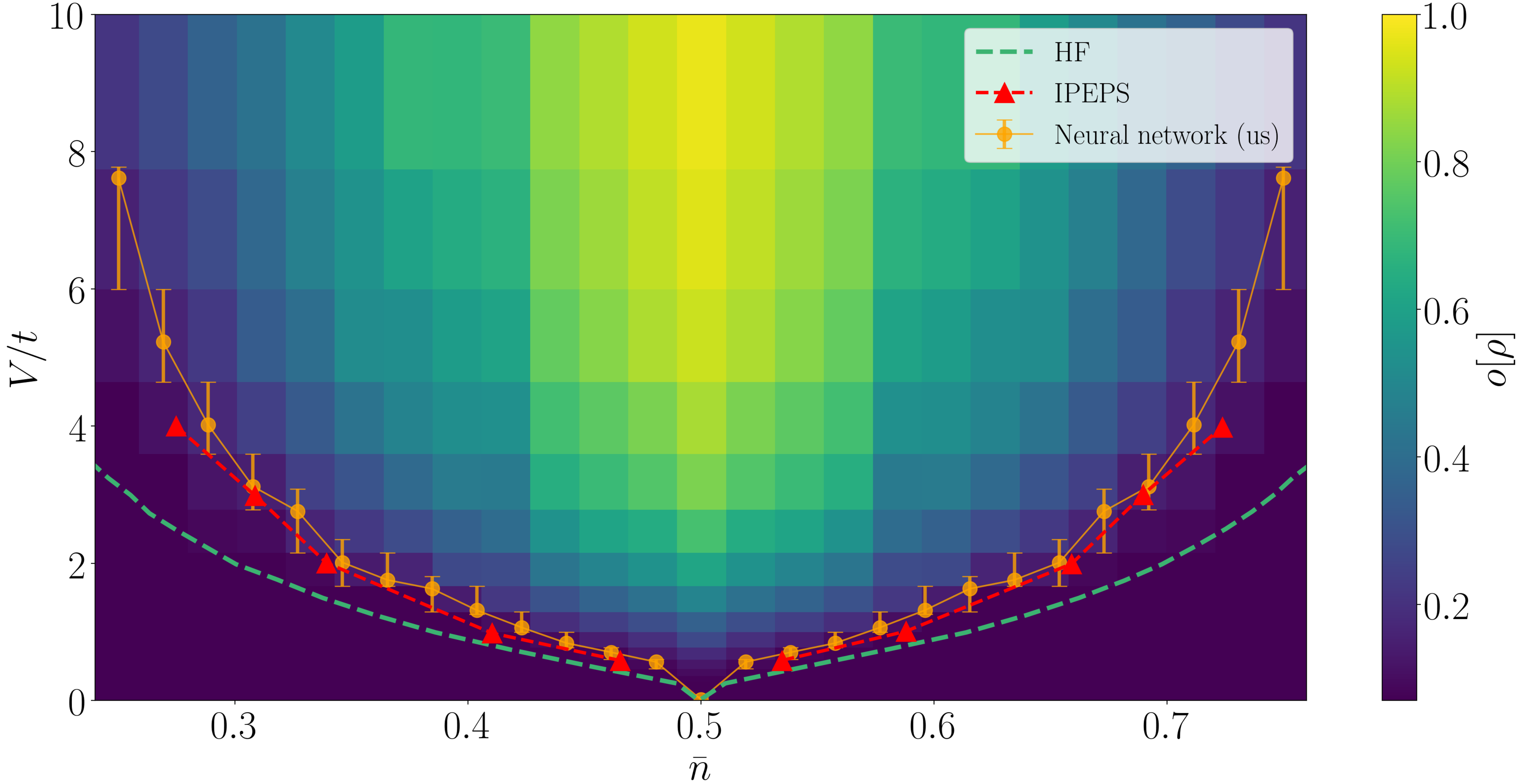}
        
        \caption{\label{FIG_06: phase diagram} Phase diagram of the two-dimensional interacting spinless fermion model under consideration. The Hartree-Fock and IPEPS transition lines are from~\cite{corboz2010IPEPS}. The orange dots correspond to the transition points from the finite-size scaling of $o[\rho]$, using the proposed variational ansatz. Lines connecting the dots are for visual guidance. The color-map represents the CDW order parameter $o[\rho]$ in the largest system size studied ($L = 10$) at different values of the particle density and interaction strength.}
\end{figure*} 
\subsection{Phase diagram}
\subsubsection{Correlation functions}
The ground-state two-point density correlation functions are shown for the largest available system size ($L = 10$) in Fig.~\ref{FIG_03: radial correlations} a). For weak coupling, the correlation functions barely oscillate and decay to zero with increasing graph distance $r$ between correlation pairs. As the interaction strength increases, the correlations spatially oscillate with increasing amplitude due to ordering of the charge distribution when the transition is crossed. In the charge-ordered phase, the amplitude of the oscillations decays (as a function of $r$) at a slower rate than in the metallic phase. At half occupation the charge ordering is staggered, leading to commensurate oscillations in the correlation functions. Away from half occupation, due to the geometry of the lattice, the charge order cannot be staggered, leading to incommensurate  oscillations in the correlations. For all occupations, increasing the interaction strength increases the amplitude of the oscillations, without significantly altering their wave form. 

This behaviour is also evident in the Fourier component amplitudes $|\widetilde{C}_\textrm{ren}(k)|$ (Fig.~\ref{FIG_03: radial correlations} b)). For all values of occupation and for weak interaction, $|\widetilde{C}_\textrm{ren}(k)|$  exhibits a uniform distribution without peaks. When the coupling is large enough, the system transitions to the charge-ordered state where a peak appears in which the amplitude monotonically increases with $V/t$. Note that at $\bar{n} = 0.24$ the peak is not well resolved due to the small number of $k$ values accessible in this system size. The position of the peak depends on the filling as anticipated. At half occupation the peak is narrow and centered around $k = \pi$ (staggered ordering). Close to half occupation and for the considered system size, the peak is still centered at $k = \pi$ but with increased width due to the rise of non-commensurate order. Lower values of the filling lead to Fourier peaks that correspond to longer wavelength orderings.

The results discussed above appear to be consistent with the formation of a CDW state as opposed to a phase-separated one. We further analyze the nature of the $|\widetilde{C}_{\rm ren}(k)|$ peak by studying its amplitude as a function of the system size, as shown in Fig.~\ref{FIG_03: radial correlations} c). At values of $V/t$ corresponding to a metallic state, the amplitude of the corresponding Fourier mode saturates to a constant value, or decreases as $L$ is increased, depending on the filling. In contrast, in the CDW phase the amplitude of the peak increases with $L$.  Although this scaling could in principle be used to determine the transition points, the system sizes we considered are not sufficient to accurately extract the critical point. This is in part due to the access to a limited set of Fourier modes in the discrete Fourier transform, which may not
provide the necessary resolution in $k$-space to resolve the true period of the oscillations away from half filling. Despite these difficulties, the above results are consistent with a CDW phase\cite{Woul2010_pahse_diag}, rather than a phase separated one.

\subsubsection{Order parameter}

 Fig.~\ref{FIG_05: density power spectrum} shows the order parameter Eq.~\eqref{Eq: order parameter} as a function of $V/t$ at different lattice occupations. Each panel also displays the order parameter at different system sizes $L = 6$, $L = 8$ and $L = 10$. Interpolation is required to obtain the value of $o[\rho]$ at the desired filling at a given system size. Linear interpolation is used instead of higher order interpolation schemes. The reason for this is that the order parameter takes values close to zero in the metallic phase, leading to negative values of the interpolated value of $o[\rho]$ when using higher order methods. The order parameter takes a small but nonzero value in the metallic phase (smaller values of $V/t$) and abruptly increases upon crossing to the charge-ordered phase. In the metallic phase and at fixed values of the filling and coupling,  $o[\rho]\geq 0$ decreases with system size. This is consistent with the CDW coming from the opening of a gap like in the $d = 1$ case, where the amplitude of the charge density wave increases with the magnitude of the gap~\cite{Giamarchi2003}. In finite-sized systems the metallic phase has a small but nonzero gap, which decreases with the increase of system size, and becomes zero in the thermodynamic limit. This nonzero gap leads to a small amplitude in the charge order. 

Finite-size scaling is thus required to find the transition points. At fixed values of the density we analyze the order parameter as a function of $V/t$, starting in the metallic phase, where its value decreases with $L$. The transition point is taken where the order parameter curves corresponding to different system sizes cross each other upon increasing the value of the coupling. Transition points are determined by the average of the first three crossing points. Error bars are determined by maximum between: the range of $V/t$ between the first three crossing points and the separation of the $V/t$ values sampled. The scaling of the order parameter at half filling is consistent with a transition at an infinitesimally small value of $V/t$ due to Fermi surface nesting~\cite{johannes2008nesting}.

The phase diagram arising from these transition points is displayed in Fig.~\ref{FIG_06: phase diagram}. Particle-hole symmetry has been applied to determine the phase boundary for $\bar{n}>0.5$ from the results obtained at $\bar{n}<0.5$.

\subsubsection{Phase diagram}

We conclude this study by analyzing the obtained phase diagram and comparing it to the phase diagram obtained with unrestricted Hartree-Fock~\cite{Czart2008HF_phase_diag} and IPEPS~\cite{corboz2010IPEPS}, in the grand canonical ensemble. Fig.~\ref{FIG_06: phase diagram} shows the phase diagrams obtained with unrestricted HF~\cite{Czart2008HF_phase_diag}, IPEPS~\cite{corboz2010IPEPS} and with our neural network ansatz, superimposed to a color-map of the order parameter in the largest system size considered. In the IPEPS study~\cite{corboz2010IPEPS} it was found that by increasing the bond dimension (and consequently the accuracy of the ground state approximation), the transition line shifts to higher values of $V/t$.
The phase boundary obtained in this work lies at slightly larger values of the interaction than those obtained from IPEPS, which may indicate that the proposed neural network anzatz is more accurate than the IPEPS wavefunction.

The magnitude of the order parameter in the largest system size analyzed $L = 10$ already provides a good indication of the transition point as shown by the superimposed color-map.

\section{Conclusion}
We showed that neural networks can be used to analyze the ground-state properties of lattice fermionic systems in first quantization. The proposed wavefunction and minimization scheme are applicable to arbitrary lattice models. In particular, we applied it to the study of the phase diagram of the two-dimensional periodically identified square lattice  with nearest neighbour repulsive interactions.

The exact diagonalization benchmarks demonstrate that the proposed wavefunction accurately captures the ground-state energy for a wide range of lattice fillings and interaction strengths. It also achieves lower energies than other approaches at half occupation for system size $L = 4$ where comparison data is available. Furthermore, we tested the accuracy in reproducing other observables such as two-point density correlation functions finding essentially the same values as those from exact diagonalization.

The study of the two-point density correlation functions shows results consistent with a charge-density-wave state for large values of the coupling, rather than a phase separated one. A finite size scaling analysis of the order parameter allowed to obtain the phase boundaries for the model, allowing the construction of the phase diagram from a canonical ensemble approach for the first time.

\section*{Acknowledgements}
JRM acknowledges support from the CCQ graduate fellowship in computational quantum physics. The Flatiron Institute is a division of the Simons Foundation. The authors acknowledge stimulating discussions with Antoine Georges, Risi Kondor, Erik Thiede and Lei Wang.

\newpage
\bibliography{references}

\onecolumngrid
\appendix

\section{Supplementary Material}
In this appendix we review the mathematical formulation of variational Monte Carlo simulation of fermionic quantum systems with a conserved fermion number $\hat{N}$. For simplicity we focus on spinless fermions with nearest-neighbor interactions.
\subsection{Review of first quantization}
Each vertex $i \in \mathcal{V}$ is associated with an operator $\hat{c}_i : \mathcal{H} \to \mathcal{H}$ acting on a fixed finite-dimensional complex Euclidean space $\mathcal{H}$. These operators satisfy the following algebra for all vertices $i,j\in\mathcal{V}$,
\begin{equation}
	\{ \hat{c}_i, \hat{c}_j^\dag \} = \delta_{ij} \, \quad \quad \{ \hat{c}_i, \hat{c}_j \} = 0
\end{equation}
where $\hat{c}_j^\dag : \mathcal{H} \to \mathcal{H}$ denotes the adjoint map of $\hat{c}_j$. Various properties of the Hilbert space can be deduced from the above algebra. In particular, there exists unit vector $| 0 \rangle \in \mathcal{H}$ such that for all $i \in \mathcal{V}$ we have $\hat{c}_i |0\rangle = 0$ (see \cite{nielsen2005fermionic} for a clean review).

Given $n \geq 0$ and a length-$n$ array $x = (i_1,\ldots, i_n) \in \mathcal{V}^N$ we introduce the following shorthand notation
\begin{equation}
	| x \rangle = | i_1,\ldots, i_n \rangle := \hat{c}_{i_1}^\dag \cdots \hat{c}_{i_n}^\dag |0\rangle.
\end{equation}

 An orthonormal basis for $\mathcal{H}$ is given by unit vectors $| i_1, \ldots, i_n \rangle$ where $\{ i_1, \ldots, i_n \} \subseteq \mathcal{V}$ denotes a subset of the vertex set of size $0 \leq n \leq |\mathcal{V}|$\footnote{In order to fix a basis, one must consider an ordering on the vertex set.}.
Since the number of subsets of $\mathcal{V}$ of size $n$ is $C^{|\mathcal{V}|}_n$ we find that the dimension of the Hilbert space suffers from the expected curse of dimensionality,
\begin{equation}
	\dim \mathcal{H} = C^{|\mathcal{V}|}_0 + C^{|\mathcal{V}|}_1 + C^{|\mathcal{V}|}_2 + \cdots + C^{|\mathcal{V}|}_{M} = 2^{|\mathcal{V}|}.
\end{equation} 
The identity operator $\mathbbm{1}$ on $\mathcal{H}$ is given by
 \begin{equation}
 	\mathbbm{1} = \sum_{0 \leq n\leq M} \frac{1}{n!} \sum_{(i_1,\ldots, i_n) \in \mathcal{V}^N} | i_1, \ldots, i_n \rangle \langle i_1, \ldots, i_n |.
 \end{equation}

It is convenient to introduce the following Hermitian operator $\hat{N} : \mathcal{H} \to \mathcal{H}$,
\begin{equation}
	\hat{N} = \sum_{i \in \mathcal{V}} \hat{c}_i^\dag \hat{c}_i .
\end{equation}
Notice that 
\begin{equation}
	\hat{N} | i_1,\ldots, i_n \rangle = n | i_1, \ldots, i_n \rangle
\end{equation}
so we have found a basis of eigenvectors of $\hat{N}$ and thus $\hat{N}$ is diagonalizable with eigenvalues $0 \leq n \leq |\mathcal{V}|$. The corresponding eigenspaces $\mathcal{H}_n$ are of dimension $C^{V}_n$ and the Hilbert space is a direct sum of eigenspaces,
\begin{equation}
	\mathcal{H} = \bigoplus_{0 \leq n \leq V} \mathcal{H}_n .
\end{equation}
The eigenvalue $n$ associated to each subspace is referred to as the particle number.

\subsection{Local energy}
Any vector $|\Psi \rangle \in \mathcal{H}_N$ can be expanded in terms of the spanning set $\{ |x \rangle : x \in \mathcal{V}^N \}$ with coefficients given by an anti-symmetric function $f : \mathcal{V}^N \to \mathbb{C}$ as follows,
\begin{align}
	\Psi
		& =  \sum_{x \in \mathcal{V}^N} f(x) \, | x \rangle \enspace ,
\end{align}
where
\begin{equation}
\langle x | \Psi \rangle = \sum_{x'\in \mathcal{V}^N} f(x') \langle x | x' \rangle = \sum_{\sigma \in S_N} f(\sigma \cdot x) \langle x | \sigma \cdot x \rangle
=
\sum_{\sigma \in S_N} f( x) \langle x | x \rangle \sgn(\sigma)^2
= N! f(x)
\end{equation}
and the squared norm of $|\Psi\rangle$ is given by
\begin{equation}
\langle \Psi | \Psi \rangle = \sum_{x,x'\in \mathcal{V}^N} f^\ast(x')f(x) \langle x' | x \rangle = N!\sum_{x \in \mathcal{V}^N} |f(x)|^2    
\end{equation}
Expanding the numerator of the Rayleigh quotient we obtain,
\begin{align}
	\langle \Psi | \hat{H} | \Psi \rangle
		& = \sum_{x \in \mathcal{V}^N} f^\ast(x) \langle x | \hat{H} | \Psi \rangle \\
		& = \sum_{x \in \mathcal{V}^N: f(x) \neq 0} f^\ast(x) \langle x | \hat{H} | \Psi \rangle \\
		& = \sum_{x \in \mathcal{V}^N : f(x) \neq 0} |f(x)|^2\left[\frac{\langle x | \hat{H} | \Psi \rangle}{f(x)}\right].
\end{align}
Dividing by the normalization gives the following expression for the Rayleigh quotient,
\begin{equation}
	\frac{\langle \Psi | \hat{H} | \Psi \rangle}{\langle \Psi | \Psi \rangle}
		= \underset{x \sim \pi_f}{\mathbb{E}}[E_{f}(x)]
\end{equation}
where
\begin{equation}
	\pi_f(x) = \frac{|f(x)|^2}{\sum_{x' \in \mathcal{V}^N}|f(x')|^2}
\end{equation}
is a probability distribution over the set $\mathcal{V}^N$ and we have defined the local energy functional
\begin{align}
    E_{f}(x) 
        & = \frac{1}{N!} \frac{\langle x | \hat{H} | \Psi \rangle}{f(x)} \enspace , \\
        & = 
        \frac{1}{N!}\sum_{x'\in \mathcal{V}^N}
        \langle x | \hat{H} | x' \rangle
        \frac{f(x')}{f(x)}
\end{align}
The Hamiltonian is of the form $\hat{H} = \hat{T} + \hat{U}$, so we consider the potential and kinetic terms separately.
Since $\hat{U}$ is diagonal, and using the antisymmetry of $f$ we obtain,
\begin{align}
	U(x)
		& :=
        \frac{1}{N!}
        \sum_{x' \in \mathcal{V}^N}
        \langle x | \hat{U} | x' \rangle
        \frac{f(x')}{f(x)}\\
        & =
        \frac{1}{N!} \sum_{\sigma \in S_N} \langle x | \hat{U} | \sigma \cdot x \rangle\frac{f(\sigma \cdot x)}{f(x)} \\
		& = \langle x | \hat{U} | x \rangle
\end{align}
Hence
\begin{equation}
	U(x) 
	= V\sum_{\{i,j\} \in \mathcal{E}} n_i n_j \enspace .
\end{equation}
Let $\Delta_x \subseteq \mathcal{V}^N$ denote the set of classical configurations obtained by applying the hopping operator $\sum_{\{i,j\}\in \mathcal{E}}(\hat{c}^\dag_i \hat{c}_j + \hat{c}^\dag_j \hat{c}_i)$ to the quantum state $|x\rangle$. Since the hopping operator is bosonic, we have
\begin{equation}
	\hat{T}|x\rangle
	=
	-t \sum_{x' \in \Delta_x} |x'\rangle
\end{equation}
and moreover since $\hat{T}$ is Hermitian,
\begin{equation}
	\langle x |\hat{T}
	=
	-t \sum_{x' \in \Delta_x} \langle x'|
\end{equation}
Thus
\begin{equation}
T(x) 
	:= \frac{1}{N!} \sum_{x'' \in \mathcal{V}^N} \langle x | \hat{T} | x'' \rangle\frac{f(x'')}{f(x)}
	= \frac{1}{N!} \sum_{x'' \in \mathcal{V}^N}\left[-t \sum_{x' \in \Delta_x} \langle x'|\right]|x''\rangle \frac{f(x'')}{f(x)}
\end{equation}
Interchanging the summations we obtain
\begin{align}
	T(x)
		& = -\frac{t}{N!}\sum_{x' \in \Delta_x} \sum_{x'' \in \mathcal{V}^N} \langle x' | x'' \rangle \frac{f(x'')}{f(x)} \\
		& = -\frac{t}{N!}\sum_{x' \in \Delta_x} \sum_{\sigma \in S_N} \langle x' | \sigma \cdot x' \rangle \frac{f(\sigma \cdot x')}{f(x)} \\
		& = - t \sum_{x' \in \Delta_x} \frac{f (x')}{f (x)} 
\end{align}
Thus,
\begin{equation}
    E_{f}(x) = -t \sum_{x' \in \Delta_x} \frac{f (x')}{f (x)} + V\sum_{\{i,j\} \in \mathcal{E}} n_i n_j
\end{equation}

\end{document}